\documentclass[12pt]{article}
\usepackage{epsfig}
\usepackage{natbib}
\usepackage{graphicx}
\usepackage{graphics}

\def\kms{\,km$\,$s$^{-1}$}
\def\deg{\hbox{$^\circ$}}
\def\sun{\hbox{$\odot$}}
\begin{document}

\noindent {\bf Investigation of the obscuring circumnuclear torus in the
active galaxy Mrk231}\\
\vspace{.1cm}

\centerline{Hans--Rainer Kl\"ockner$^{1,2}$, Willem A. Baan$^{2}$ 
and Michael A. Garrett$^{3}$ \\}

\vspace{0.3cm}

\noindent{\footnotesize 1. Kapteyn Astronomical Institute, University of Groningen, 9700 AV, Groningen, The Netherlands\\
2. ASTRON, P.O.Box 2, 7990 AA, Dwingeloo, The Netherlands\\
3. JIVE, Joint Institute for VLBI in Europe, PO Box 2, 7990 AA Dwingeloo, The Netherlands\\}

\noindent{published in Nature Vol. 421, 821--823, 20 February 2003\\}

\vspace{0.2cm}

\noindent{\bf \sf Active galaxies are characterized by prominent
emission from their nuclei. In the unified view of active galaxies,
the accretion of material onto a massive compact object - now
generally believed to be a black hole - provides the fundamental power
source \citep{1984ARAA..22..471R}. Obscuring material along the line
of sight can account for the observed differences in nuclear emission
\citep{1988ApJ...329..702K, 1999agnc.book.....K}, which determine the
classification of AGN (for example, as Seyfert 1 or Seyfert 2
galaxies). Although the physical processes of accretion have been confirmed
observationally \citep{1995ApJ...440..619G,1997Natur.388..852G}, the
structure and extent of the obscuring material have not been
determined. Here we report observations of powerful hydroxyl (OH) line
emissions that trace this obscuring material within the circumnuclear
environment of the galaxy Markarian~231. The hydroxyl (mega)--maser
emission shows the characteristics of a rotating, dusty, molecular
torus (or thick disk) located between 30 and 100~pc from the central
engine. We now have a clear view of the physical conditions, the
kinematics and the spatial structure of this material on intermediate
size scales, confirming the main tenets of unification models.}

\vspace{0.2cm}

\noindent{Observations of powerful molecular water--vapour and hydroxyl
megamaser emissions , which are a million times more luminous than
their galactic counterparts \citep{1985Natur.315...26B}, have provided
unique information about distinct regions within galactic
nuclei. Water--vapour masers can map out the central few parsecs of an
AGN, tracing the rotation of nuclear accretion disks and the state of
molecular material surrounding the radio jets
\citep{2002PASA...19..401M}. The inferred geometry and thickness of
these disks, however, cannot account for the observed obscuration
signature and the radiation patterns seen in active galactic nuclei
(AGN) \citep{2002PASA...19..401M, 1998ApJ...508..243H}. On the other
hand, the hydroxyl masers in galactic nuclei are spread across
hundreds of parsecs
\citep{1999ApJ...511..178D,1998ApJ...493L..13L,2001AA...377..413P}. Their
tight correlation with the infrared emission
\citep{1989ApJ...338..804B} indicates a strong association with the
dusty obscuring material along the line of sight towards the nucleus
\citep{1988ApJ...329..702K,1999agnc.book.....K}. To study the
distribution of the hydroxyl megamaser emission with respect to the
nuclear engine at a resolution of tens of parsecs in Mrk~231, we
performed very--long--baseline interferometry (VLBI) radio
observations with the European VLBI Network (EVN). Mrk~231 is the most
luminous infrared galaxy in the local universe (distance of 172 Mpc),
and is undergoing a merger as indicated by the high dust content and
the prominent tidal tails observed at optical wavelengths. The nuclear
power source has been classified \citep{1998ApJ...509..633B} as a
Seyfert~1, which in the unification scheme would suggest that the
observer has a direct, and almost unobscured, line--of--sight view into
the broad--line emission region at parsec scales. The location of the
nuclear engine on similar scale sizes is traced by the radio outflow
\citep{1999ApJ...517L..81U}, which extends to almost a hundred parsecs
\citep{1999ApJ...516..127U}.}

\vspace{0.2cm}

\begin{figure}
\includegraphics[angle=0,width=12cm]{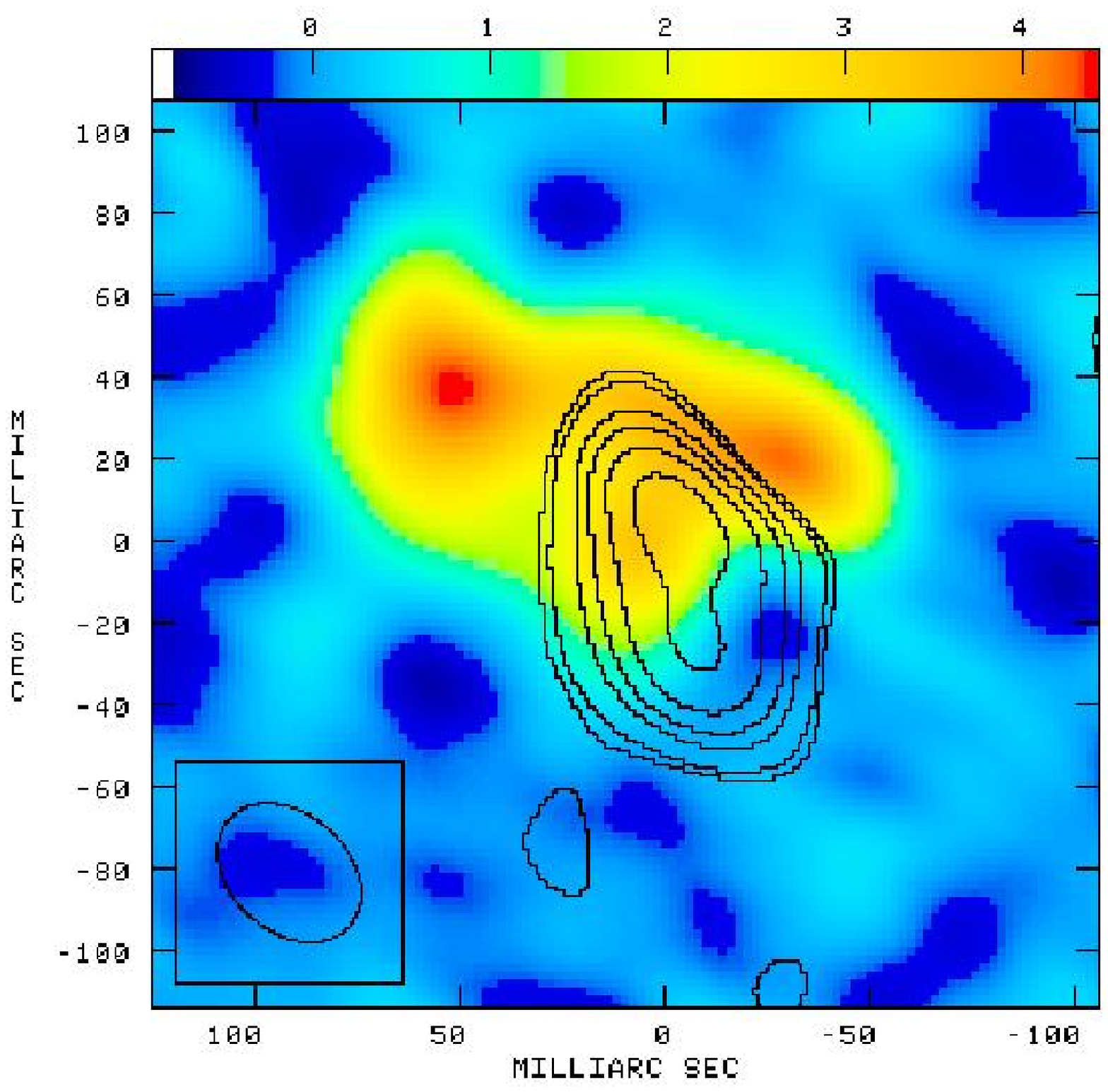}
\caption{{\bf Hydroxyl and radio--continuum emission in the
nucleus of Mrk 231.}
\newline
The integrated OH--line emission in pseudo--color (in mJy per beam) is
superimposed on the nuclear continuum emission in contours (peak 39
mJy per beam). The contour levels are a geometric progression in
factors of the square root of 2 starting at 4 mJy per beam. These EVN
observations of Mrk~231 were made on September 1999 for 12 hrs in
dual--polarization phase--referencing mode at 1599 MHz and with 256
spectral channels across a total bandwidth of 8 MHz. The synthesized
beam is 39 mas $\times$ 28 mas, where 1~mas corresponds to a size of
0.83 pc at the distance of Mrk 231 (172 Mpc by assuming q$_0$=0.5 \kms\
and H$_0$=75 \kms\ Mpc$^{-1}$). The data has been correlated at the EVN
correlator at the Joint Institute for VLBI in Europe (JIVE), and
later transferred to the AIPS packages for calibration of the complex
visibilities. The integrated line emission image has been obtained by
averaging over the velocity range of the OH lines, after subtraction
of the radio continuum, obtained by averaging the channels without 
line emission.}
\label{fig:mrk231evncontoh}
\end{figure}

\begin{figure}
\hspace{2.5cm}
\includegraphics[angle=0,width=8cm]{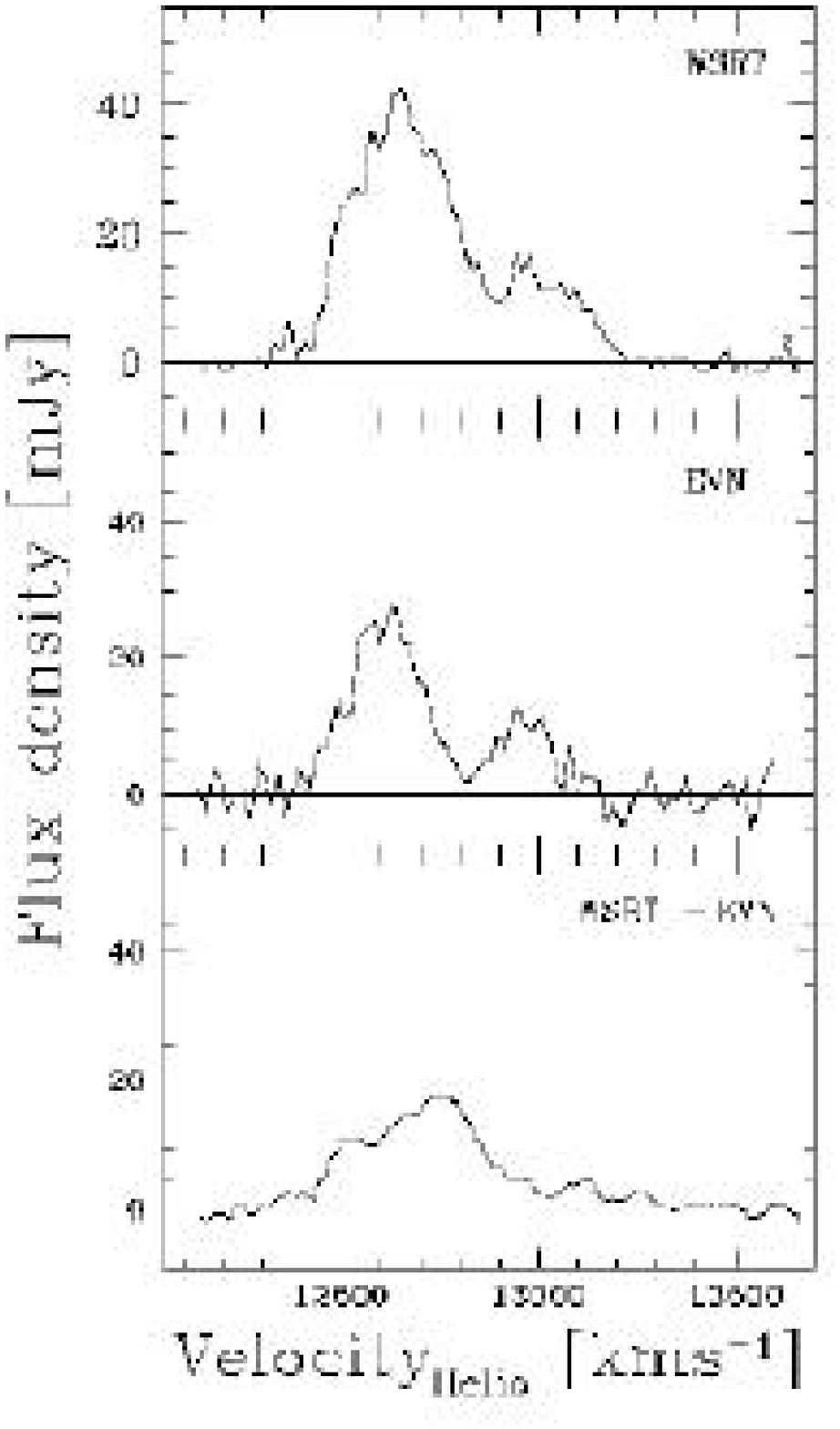}
\caption{{\bf Hydroxyl emission spectra of Mrk 231 traced at
different scale sizes.}
\newline
The top panel shows the WSRT spectrum taken at 14 arcsec resolution,
the middle panel displays the spectrum for the EVN observations at 39
mas resolution, and the bottom panel displays the difference spectrum
of both observations. The emission line features show both OH main
lines at 1667 and 1665~MHz that are offset by 365~\kms ; the velocity
scale refers to the 1667~MHz emission line by assuming a heliocentric
velocity of V$_{\rm center}$ = 12650 \kms\ based on the CO(1-0) centre
velocity \citep{1998ApJ...507..615D}. The spectral resolution is
18~\kms\ per channel. The OH line width in the EVN data is 214~$\pm$~11
\kms\ (FWHM), which is 70~\kms\ less than for the WSRT data.  The
residual emission line in the bottom frame counts for 44 \% of
the OH emission in Mrk 231 and is systematically red--shifted by around
45~\kms. Similar flux discrepancies, but with no distinct velocity
offset, have been found for other OH--MM galaxies studied at high
resolution
\citep{1999ApJ...511..178D,1998ApJ...493L..13L,2001AA...377..413P}. Similar
velocity offsets have also been found for the CO(2-1) emission
\citep{1996ApJ...457..678B} and the HI absorption in Mrk 231
\citep{1998AJ....115..928C}.}
\label{fig:mrk231spectra}
\end{figure}

\noindent{The EVN data reveal an OH emission structure with an
east--west extent of 130~pc that is straddling a north--south
elongated radio continuum emission (Fig.~\ref{fig:mrk231evncontoh}). This
continuum emission has been further resolved into a lobe--core--lobe
structure \citep{1999ApJ...516..127U} of 80~pc at position angle PA =
8\deg\ , which is found to be misaligned by 57\deg\ with the core
radio structure on parsec scales \citep{1999ApJ...517L..81U}. The
emission spectrum integrated across the whole OH emitting region
displays the two main hydroxyl lines at 1667 and 1665 MHz showing a
line--width of around 200 \kms\ (full--width at half--maximum, FWHM) at
a centroid velocity of 12610~\kms (Fig.~\ref{fig:mrk231spectra}). The EVN
data and also the intermediate--scale MERLIN
\citep{2000IAUS..205E..27R} observations account for only half of the
integrated line emission seen by the Westerbork Synthesis Radio
Telescope (WSRT; see Figure 2) and previous single--dish measurements
\citep{1985Natur.315...26B}. This missing emission component must
therefore originate within the resolution gap between the EVN and WSRT
observations, in an extended region between 190~pc and 11~kpc of the
nucleus. This component is probably associated with the nuclear
continuum emission \citep{1999ApJ...519..185T} and the CO(1-0) line
emission \citep{1996ApJ...457..678B, 1998ApJ...507..615D}. Figure 3
(top panel) presents the spatial velocity distribution, which is
similar for each of the two hydroxyl lines and which has an overall
velocity shift of about 160~\kms\ from the northwest towards the
southeast. The velocity width of the emission lines has an average
value of about 70~\kms\ (FWHM) across the northern part of the radio
continuum emission. On the other hand, there is also a distinct OH
emission region at the northwestern edge of the continuum emission
that has a velocity similar to that seen towards the centre of the
radio core, and which has a distinctly higher velocity width of
85~\kms\ (Figure 3; middle panel).}

\vspace{0.2cm}

\noindent{The nature of the OH emission observed in the nucleus of
Mrk~231 adds a new component to the hierarchical structure of this
system, and complements the information obtained with other tracers at
lower resolution. We find that the OH velocity field with its symmetry
axis at PA = 34\deg does not agree with that of the kiloparsec--scale
CO(1-0) structure \citep{1996ApJ...457..678B, 1998ApJ...507..615D}
having a PA = 0\deg . Neither does it agree with the large--scale radio
outflow (PA = 8\deg), or the more compact CO(2-1)
distribution \citep{1998ApJ...507..615D} extending up to 850~pc at PA =
18\deg . However, the OH velocity field is in agreement with
that of the line--of--sight HI absorption \citep{1998AJ....115..928C}
at PA = 27\deg . The HI absorption is detected against a diffuse
radio--emitting halo with an extent of about 350~pc and an orientation
of PA = 22\deg . This diffuse radio halo, which is possibly associated
with a circumnuclear starburst \citep{1999ApJ...519..185T}, could then
also trace the dense and opaque interstellar medium of a disk or
torus \citep{1998AJ....115..928C} inclined at 56\deg . The integrated
HI velocity dispersion \citep{1998AJ....115..928C} of 193~$\pm$~25 \kms\
agrees well with that of the hydroxyl lines
(Fig.~\ref{fig:mrk231spectra}). This suggests that the OH emission indeed
originates in the central regions of a circumnuclear rotating disk or
torus having an outer structure that is traced by the CO(2-1)
emission.}

\vspace{0.2cm}

\noindent{The energy source responsible for the excitation of the OH
molecules is the extreme far--infrared radiation
\citep{1989ApJ...338..804B} field in Mrk~231. The observed
far--infrared spectral temperature of around 43 K provides the ideal
pumping conditions for the OH clouds, which could range in size from
10~pc up to 100~pc \citep{1987AA...185...14H,1995AA...300..659R}. The
ratio of the intensity of the two maser lines varies rather smoothly
across the entire emission structure and has an average value of 1.8,
as expected for local temperature equilibrium (LTE). This would
indicate that the observed emission lines are optically thin and are
mostly unsaturated masers. Assuming that the diffuse radio continuum,
which serves as a background for the nuclear HI absorption, also
serves as a background for the observed OH masers, this emission could
be explained in terms of exponential amplification with a maximal gain
of about 2.2 within classical OH maser--models
\citep{1982ApJ...260L..49B,1985Natur.315...26B}. However, extreme
values for the line ratio have been found in the discrepant
northwestern region discussed earlier in this text. This region must
have a completely different pumping environment than the rest of the
structure. The unusual blue--shifted velocity, the slightly higher
velocity dispersion, and the line ratios of this discrepant region are
all consistent with the theoretically predictions
\citep{1985Natur.315...26B,1995AA...300..659R,2002ApJ...566L..21W,1985Natur.315...26B} for an interaction region between the radio outflow and the molecular
environment.}

\vspace{0.2cm}

\noindent{The molecular environment described above for the nuclear
region of Mrk~231 is very similar to the dust structures found in
active galaxies such as NGC~4261 using the Hubble Space Telescope
\citep{2000AA...354L..45V}. We have applied an axisymmetric torus
model with a circular cross--section to the EVN data, in order to
determine the structural boundaries in the nuclear region. The compact
molecular structure surrounding the nucleus is misaligned with the
large--scale molecular disk of the galaxy by 34\deg . The radius of
the inner cavity of the torus would be about 30~pc, as suggested by
the location of the northwestern interaction region. The outer edge of
the torus spans a diameter of 200~pc, which is based on the 4~mJy
contours of the diffuse radio emission \citep{1998AJ....115..928C} and
on far--infrared blackbody estimates \citep{1998ApJ...507..615D}. This
region also serves as a radio background for the OH amplification
process. These parameters result in a torus structure with a radius of
65~pc and a thickness of 70~pc, which accounts for an obscuration
angle of 60\deg\ centered on the plane of the torus and which is
representative of values found for other galaxies
\citep{2001ApJ...555..663S}. The orientation and the shape of the
extended radio contours and the symmetry axis of the velocity field
indicate that the torus is tilted upward from the line of sight by
56\deg , and is rotated anticlockwise by 35\deg . This inferred
model for the torus has been displayed as a wire--structure
representation and is presented in Figure 3 (bottom panel). The
nuclear radiation cones with opening angles of 60\deg\ , assumed to be
similar to those seen in M~87 \citep{1999Natur.401..891J}, represent the
directions with an unobscured view of the nucleus. The radio outflow
follows a twisted path within these cones, starting at the nucleus (at
PA = 65\deg) and ending up as a double source aligned with the
symmetry axis of the outer disk (at PA = 8\deg)
\citep{1999ApJ...517L..81U, 1999ApJ...516..127U}. As a result, this
parameterization indicates that the nuclear accretion disk, that
collimates the jet outflow inside the central cavity, is actually
misaligned by 30\deg\ with the plane of the molecular torus.}

\vspace{0.2cm}

\noindent{In reality, the proposed torus structure does not need to
have a circular cross--section. In addition, it will most probably be
wrapped inside a cocoon--like surface region with a higher
temperature, and will have the same outer extent of 460~pc as the
diffuse radio emission and the HI absorbing
gas \citep{1998AJ....115..928C}. However, the representation of the
inner part of this rotating dusty and molecular torus in Mrk~231 is
consistent with (all) current models of galactic nuclei and
theoretical investigations \citep{1999agnc.book.....K}, and supports
the unification schemes for AGN. Whether a different orientation of
the torus and the nuclear accretion disk represents a special case or
is a general characteristic of active nuclei needs to be investigated
with VLBI observations of other megamaser galaxies hosting a similar
nuclear power source.
}
\vspace {0.2cm}

\noindent{\bf Acknowledgments}
\vspace {0.1cm}

\noindent{We thank C. Carilli for providing a map of the diffuse continuum
structure in Mrk~231. H.-R.K thanks O. M\"oller for advice on
programming in OpenGL software. The European VLBI Network is a joint
facility of European, South African, and Chinese radio astronomy
institutes funded by their national research councils. The Westerbork
Synthesis Radio Telescope is operated by the ASTRON (Netherlands
Foundation for Research in Astronomy) with support from the
Netherlands Foundation for Scientific Research NWO.\\}

\vspace {0.2cm}

\noindent Correspondence and requests for material should be addressed to H.--R. K.(e--mail: hrkloeck@astro.rug.nl).

%
%

\newpage
\pagestyle{empty}

\newpage
\pagestyle{empty}

\newpage
\pagestyle{empty}
\begin{figure}
\hspace{4cm}
\includegraphics[angle=0,width=4cm]{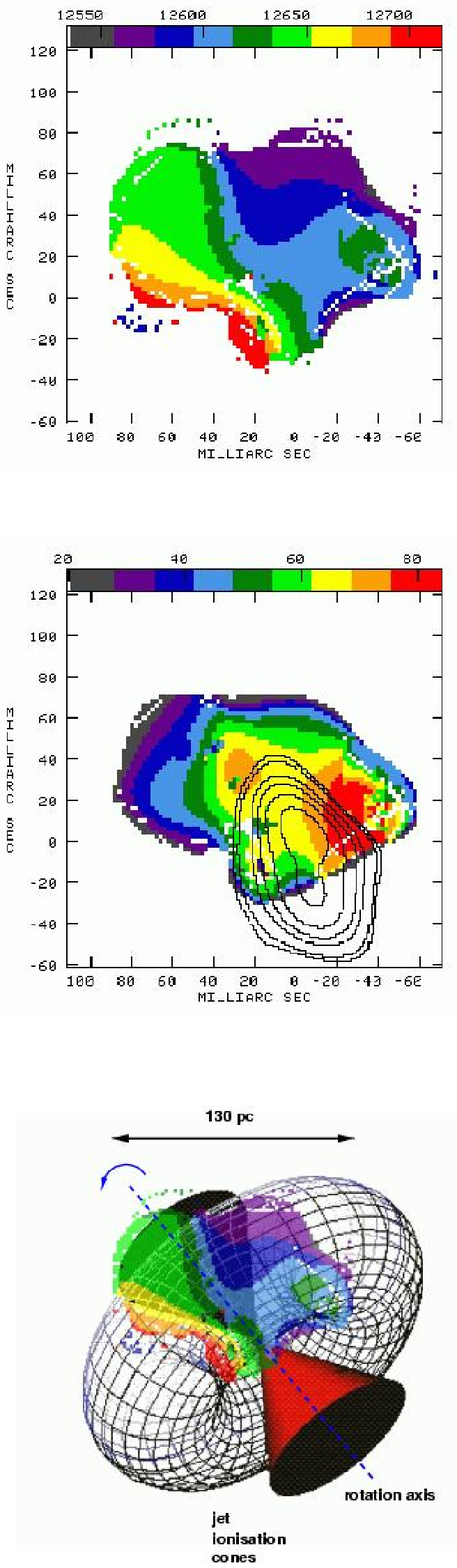}
\caption{{\bf The circum--nuclear kinematics in Mrk 231.}
\newline
\underline{Top panel} - The velocity field of the OH 1667 MHz emission in the EVN data
is shown in pseudo--colors. The spatial resolution is the same as for
Fig.~\ref{fig:mrk231evncontoh} and the velocity scale is the same as in
Fig.~\ref{fig:mrk231spectra}. The overall velocity pattern reveals an
east--west velocity gradient of 1.41 \kms\ per parsec over the entire OH
emission such that the northwestern edge of the torus moves towards
the observer.  
\newline
\underline{Middle panel} - The line width of the 1667 MHz emission
lines (in \kms) is superimposed on the nuclear radio continuum
emission in contours displayed in a similar way as in Fig.~\ref{fig:mrk231evncontoh}. The
diagram shows that the velocity dispersion varies significantly
across the emission region. The velocity dispersion structure shows
several distinct regions but with the highest values to be found at
the western edge of the radio continuum source.  
\newline
\underline{Bottom panel} - The inferred model of the nuclear torus in Mrk~231 is
displayed as a wire diagram with symmetric ionization cones. This
model takes into account all large--scale characteristics of the
nuclear radio emission and the OH emission. The symmetry axis of the
torus has been inclined upwards by 56\deg , and then rotated anticlockwise by 35\deg . The molecular material moves from
top--right to bottom--left (northwest to southeast). The virial
estimates of the central mass concentration is 7.2~$\pm$~3.8
$\times$10$^7$ M$_{\sun}$.}
\label{fig:mrk231model}
\end{figure}

\end{document}